\documentstyle[12pt]{article}
\begin{document}
\def\overlay#1#2{\setbox0=\hbox{#1}\setbox1=\hbox to \wd0{\hss
#2\hss}#1\hskip -2\wd0\copy1}
\def\lsim{\mathrel{\rlap{\lower4pt\hbox{\hskip1pt$\sim$}}
    \raise1pt\hbox{$<$}}}         
\def\gsim{\mathrel{\rlap{\lower4pt\hbox{\hskip1pt$\sim$}}
    \raise1pt\hbox{$>$}}}         
\renewcommand{\thefootnote}{\fnsymbol{footnote}}
\setcounter{footnote}{1}
\baselineskip20pt
\begin{titlepage}

\rightline{Adelaide Univ. Preprint: ADP-93-211/T129}
\vskip15pt
\centerline{\bf Charge-Symmetry Breaking, Rho-Omega Mixing, and}
\centerline{\bf the Quark Propagator}

\vskip15pt
\centerline{ G.~Krein$^{a}$, A.W.~Thomas$^{b}$
and A.G.~Williams$^{b,c}$}

\vskip10pt
\centerline{$^a$Instituto de F\'{\i}sica T\'eorica, UNESP}
\centerline{Rua Pamplona 145, 01405 S\~ao Paulo, SP, Brazil}
\vskip0.3cm
\centerline{$^b$Department of Physics and Mathematical Physics,
University of Adelaide}
\centerline{GPO Box 498 Adelaide, S. Aust. 5001, Australia}
\vskip0.3cm
\centerline{$^c$Department of Physics and the Supercomputer
Computations Research Institute}
\centerline{Florida State University, Tallahassee,FL 32306, USA}
\vskip20pt

\centerline{{\bf Abstract}}

The momentum-dependence of the $\rho^0$-$\omega$ mixing
contribution to charge-symmetry breaking (CSB) in the nucleon-nucleon
interaction is compared in a variety of models.  We focus in particular on
the role that the structure of the quark propagator plays in the
predicted behaviour of the $\rho^0$-$\omega$ mixing amplitude.
We present new results for
a confining (entire) quark propagator
and for typical propagators arising from explicit numerical
solutions of quark Dyson-Schwinger equations.  We compare these to
hadronic and free quark calculations.   The implications
for our current understanding of CSB experiments is discussed.

\vfill
\noindent Manuscript available as a {\LaTeX} file
\vfill
\leftline{e-mail addresses:}
\leftline{\hskip1cm gkrein@ift.uesp.ansp.br}
\leftline{\hskip1cm athomas@physics.adelaide.edu.au}
\leftline{\hskip1cm awilliam@physics.adelaide.edu.au}
\end{titlepage}
\vfill\eject

\renewcommand{\thefootnote}{\arabic{footnote}}
\setcounter{footnote}{0}

A wide variety of charge symmetry breaking (CSB) phenomena in nuclear physics 
appear to be well explained in terms of 
one-boson exchange potentials, when electromagnetic effects,
the neutron-proton mass difference, and isoscalar-isovector meson mixing
are included. A recent and extensive review of CSB has recently
appeared\cite{mns}.  Ultimately we would like to understand CSB  
in terms of electromagnetic effects and the $u$-$d$ quark mass difference
in a microscopic quantum chromodynamics (QCD) based description of the
strong interactions.
Any potential which gives rise to CSB in the nucleon-nucleon ($NN$)
sytem can be categorized as either
a class III or class IV interaction in the conventions of
Henley and Miller\cite{hm}.

In the relatively successful explanation of a number of observed CSB
phenomena, calculated contributions from $\rho^0$-$\omega$
mixing play a major role.  These phenomena include: (i) differences
between $pp$
and $nn$ scattering lengths and effective range parameters, (ii)
binding energy difference of $^3\rm H$ and $^3\rm He$, (iii) analyzing power
difference in $n$-$p$ elastic scattering, and (iv) the Nolen-Schiffer
anomaly.
In the one-boson exchange model of the nucleon-nucleon interaction
the $\rho^0$-$\omega$ potential is generated by $\rho^0$-$\omega$ mixing
in the intermediate vector-meson 
propagators\cite{cb}.  Another source of CSB arising from meson mixing is
that from $\pi^0$-$\eta$ mixing,
but this is typically
much less important than that due to the $\rho^0$ and $\omega$.
The long standing problem of 
explaining the binding energy differences of mirror nuclei, 
(the Nolen-Schiffer anomaly)\cite{ns}, appears to have been largely
resolved as a result of the $\rho^0$-$\omega$ mixing 
potential\cite{bi}. The theoretical understanding of the difficult
neutron-proton analyzing power difference measurements carried out
at TRIUMF\cite{triumf} and  IUCF\cite{iucf} 
relied on $\rho^0$-$\omega$ mixing
and in particular this was the dominant contribution in the latter.
\cite{{hht},{mtw},{bw}}.

Although the situation appears to be very satisfactory, there is however at 
least one potential problem with the standard evaluations which has been 
pointed out recently by Goldman, Henderson and Thomas(GHT)\cite{ght}. The 
problem is associated with the {\it assumed} off-shell behavior of the
$\rho^0$-$\omega$ mixing matrix element.  The standard assumption  
is that the mixing amplitude is
momentum independent with its value extracted from the experimental data
on $e^{+}e^{-} \rightarrow \pi^{+}\pi^{-}$ at the $\omega$ pole, 
$q^2 \simeq m_{\omega}^2$, while the exchanged mesons in Eq.~(\ref{Vq}) 
have space-like four-momentum, $q^2 < 0$, and are therefore highly virtual. 
In their initial study a simple model was used where
the vector mesons are considered as 
quark-antiquark composites. Because of the the up-down quark mass difference 
the $\rho^0$-$\omega$ mixing amplitude is generated by an intermediate quark 
loop. A $\rho^0$ of four momentum
$q^{\mu}$ dissociates into a quark-antiquark pair 
and then recombines to form an $\omega$, or vice versa.
The quark and antiquark momenta are
$\ell^\mu+(q^\mu/2)$ and $-\ell^\mu+(q^\mu/2)$ respectively, where
$\ell$ is the loop momentum.
There will also be a contribution from electromagnetic effects, but
this will be much smaller than that from the quark mass difference.
Simple vertex functions were introduced and free quark propagators were 
used. Once the mixing amplitude was calculated, GHT 
obtained the coordinate space (static) potential by Fourier transforming 
the resulting momentum-space potential. The basic conclusions of the GHT
calculation  were: (i) the 
mixing amplitude is strongly momentum dependent, (ii) there is a node in the 
potential at about $0.9$fm and, (iii) because of the node, the potential 
changes sign and so its importance is greatly suppressed.

The potential implications of this result of GHT are profound
in view of the central role
meson mixing plays in our understanding of charge symmetry breaking.
If these results can be confirmed in other more realistic calculations,
it may be necessary to find new sources of CSB in order to explain
existing data. 
Since the initial investigation by GHT other studies appear to support
their basic conclusions.  These include investigations of $\pi^0$-$\eta$
mixing using a quark loop model\cite{mg}, chiral perturbation
theory\cite{km}, and a hadronic model\cite{jp}.
In addition, each of two further studies of $\rho^0$-$\omega$ mixing
using a hadronic model\cite{pw} and using QCD sum rules\cite{hhmk}
also support the basic conclusions of GHT.

In the present study we investigate the role of quark confinement and the
nature of the quark propagator in a $q$-$\bar q$ based description of
the $\rho^0$-$\omega$ mixing amplitude.
This study is motivated by one of the difficulties associated with the
GHT calculation, namely that, as their work did not include quark
confinement, an unphysical $q\bar q$-pair production threshold resulted
in the timelike region - at $q^2=4m_q^2$.  This corresponds to the vector
meson being able to dissociate into an on-shell quark-antiquark pair
for $q^2\ge 4m_q^2$.  We review some necessary formalism before
discussing details of our quark propagators. 

The interaction Lagrangian densities for the nucleon-$\rho$
and nucleon-$\omega$ couplings are
\begin{equation}
{\cal L}_{{\scriptscriptstyle NN}\rho}=
g_{\rho}\bar{\psi}\gamma^{\mu}
{\vec{\tau}}\cdot\psi{\vec{\rho}}_{\mu} +
f_{\rho}\bar{\psi}\sigma^{\mu\nu}
{\vec{\tau}}\cdot\psi{\partial_{\mu} \over 2M}{\vec{\rho}}_{\nu} \;,
\label{lrho}
\end{equation}        
\begin{equation}
{\cal L}_{{\scriptscriptstyle NN}\omega}=
g_{\omega}\bar{\psi}\gamma^{\mu}\psi{\omega}_{\mu} \;.
\label{lomega}
\end{equation}
These definitions are standard, (although some authors vary in their
definitions of the couplings by including additional factors of 1/2).
No tensor coupling for the $NN\omega$ vertex is included since
we choose typical coupling constants and form factors 
determined by the Bonn group~\cite{mhe}.
The contribution from $\rho^0$-$\omega$ mixing to
the $NN$ potential, is given by (with 
$\Gamma^{\mu}\equiv~i\sigma^{\scriptscriptstyle \mu\nu}
{q_{\nu}/2M_N}$ and $C_{\rho}\equiv f_{\rho}/g_{\rho}$)
\begin{eqnarray}
\hat{V}^{\rho\omega}_{\scriptscriptstyle III}(q) &=&
V(q) \,
\gamma^{\mu}(1)\gamma_{\mu}(2)
\Big[\tau_{\scriptscriptstyle z}(1)+\tau_{\scriptscriptstyle z}(2)\Big] \;, 
\label{vthree} \\
\hat{V}^{\rho\omega}_{\scriptscriptstyle IV}(q) &=&
V(q) \,
C_{\rho}
\Big[
\Gamma^{\mu}(1)\gamma_{\mu}(2)
\tau_{\scriptscriptstyle z}(1)-
\gamma^{\mu}(1)\Gamma_{\mu}(2)
\tau_{\scriptscriptstyle z}(2)
\Big]  \;,
\label{vfour} 
\end{eqnarray}
where
\begin{equation}
V(q)=
-{g_{\rho}g_{\omega} F_{N\rho}(q^2) F_{N\omega}(q^2)
\langle \rho^0 | H | \,\omega \rangle \over
(q^{2}-m_{\rho}^{2})
(q^{2}-m_{\omega}^{2})}\;,
\label{Vq}
\end{equation}
where $\langle\omega|H|\rho^0\rangle$
is the meson mixing amplitude as defined in Ref.~\cite{cb}.
We have introduced the standard Bonn form factors at the
nucleon-meson vertices, which for on-shell nucleons are given by
$F_{N\rho}(q^2) \equiv 1/[1-(q^2/\Lambda_\rho^2)]$
and 
$F_{\omega}(q^2) \equiv 1/[1- (q^2/\Lambda_\omega^2)]$.
With this normalization for the form factors, the appropriate
Bonn couplings are $g_\rho^2/4\pi=0.41$
and $g_\omega^2/4\pi=10.6$.  Also, $\Lambda_\rho=1400$~MeV,
$\Lambda_\omega=1500$~MeV, $C_\rho\equiv f_\rho/g_\rho=6.1$
and as already stated $C_\omega\equiv f_\omega/g_\omega=0$.

Our starting point for the present calculations is
the expression for the $\rho\omega$ mixing
self-energy in terms of a quark-antiquark loop insertion
\begin{equation}
\Pi^{\mu \nu}(q^2)=3ig_{q\rho}g_{q\omega}\int {d^4\ell \over (2\pi)^4}\; 
{\rm tr}\Big[ \Gamma^{\mu}_{\rho}(\ell_+,\ell_-) S_{u}(l_{-})
\Gamma^{\nu}_{\omega}(\ell_-,\ell_+)S_u(\ell_+)\Big]
-\Big[u\longrightarrow d\Big]\;.
\label{Pimunu}
\end{equation}
The trace is over spinor indices and the factor of three comes
having performed the colour trace.  $\Gamma_{\rho}^{\mu}(p',p)$ and
$\Gamma_{\omega}^{\mu}(p',p)$ are 
the quark-vector-meson vertex functions for the $\rho^0$ and $\omega$
respectively, $S_{i}(q)$ is the quark propagator 
with flavour $i$ ($i=u,d$) and $\ell_{\pm}\equiv \ell \pm(q/2)$.
We take for simplicity here $\Gamma_{\rho}^{\mu}(p',p)= \gamma^\mu
F_{q\rho}(p',p)$ and $\Gamma_{\omega}^{\mu}(p',p)= \gamma^\mu
F_{q\omega}(p',p)$, where $F_{q\rho}$ and $F_{q\omega}$ are the
quark-vector-meson form factors.  Since we wish to study the $q^2$-dependence
arising from the quark propagator in the loop, we follow Ref.~\cite{ght}
and consider these form factors to be functions of $\ell^2$ only, i.e.,
$F_{q\rho}(\ell_+,\ell_-)=F_{q\rho}(\ell^2)$ and 
$F_{q\omega}(\ell_-,\ell_+)=F_{q\omega}(\ell^2)$.  We do not wish to introduce
any unconstrained $q^2$-dependent behaviour through the quark-meson
vertex functions.

We can make use of Lorentz invariance and current conservation (i.e.,
$q_\mu\Pi^{\mu\nu}=0$) to write
$\Pi^{\mu\nu} \equiv \big[-g^{\mu\nu} + (q^\mu q^\nu/q^2)\big]\Pi(q^2)$,
where this is the definition of the scalar function $\Pi(q^2)$.  Note that in
the case of the
photon self-energy we would require $\Pi(0)=0$ in order that the photon
not acquire a mass through self-energy insertions, however there is no such
restriction in the present case.  Since we are only interested in coupling
to conserved external currents, (i.e., since $q_\mu J^\mu=0$ for the
nucleons), we need only
retain the $g^{\mu\nu}\Pi(q^2)$ part of Eq.~(\ref{Pimunu}) in our
calculation of the $NN$-interaction.
A comparison of the definition \cite{cb} of the mixing amplitude
$\langle\omega|H|\rho^0\rangle$ with the usual Feyman rules shows that
this is identical with the $\rho\omega$ mixing self-energy $\Pi(q^2)$,
i.e., $\langle\omega|H|\rho^0\rangle\equiv \Pi(q^2)$.

The general form of the quark propagator in Min\-kows\-ki spa\-ce is
\begin{equation}
S(q)\equiv {1\over{A(q^2)\not\!q - B(q^2)}}
\equiv {Z(q^2)\left[{\not\!q+M(q^2)}\right]
\over{q^2-M^2(q^2)}}
\equiv F(q^2)\left[{\not\!q+M(q^2)}\right]\;,
\label{Sq}
\end{equation}
where $A(q^2)$ and $B(q^2)$ are scalar functions of $q^2$ and, 
$Z(q^2)=1/A(q^2)$ and $M(q^2)=B(q^2)/A(q^2)$. 
One possible mechanism of quark confinement is that
the quark propagator does not have a mass
pole, i.e., that the function $F(q^2)\equiv~Z(q^2)/[q^2-M^2(q^2)]$
does not have a pole or is an entire function in the complex $q^2$
plane\cite{rwk}.
One sees that there is no mass pole when the function 
$Z(q^2)$ goes to zero for $q^2\longrightarrow M^2(q^2)$. One explicit
quark model where this occurs can be found in Ref~(\cite{brw})
where this property arises
from the solution of a model quark Dyson-Schwinger equation.

We can isolate the $g^{\mu\nu}\Pi(q^2)$ part of $\Pi^{\mu\nu}$ 
and we then find in our $q\bar q$-model that the mixing amplitude can be
written as
\begin{eqnarray}
&&\Pi(q^2)=3ig_{q\rho}g_{q\omega}\int {d^4\ell \over (2\pi)^4}\; 
F_{q\rho}(\ell^2)F_{q\omega}(\ell^2)
\Bigg\{\Bigg[F_u(\ell_-^2)F_u(\ell_+^2) \nonumber \\
&&\Big[\Big({8\over 3}\Big)\Big(\ell^2-
{(q\cdot\ell)^2\over q^2}\Big)-4\ell^2+q^2+4M_u(\ell_-^2)M_u(\ell_+^2)\Big]
\Bigg] - \Bigg[u\longrightarrow d\Bigg]
\Bigg\}\;,
\label{Pi}
\end{eqnarray}
where for the $u$-quark propagator $S_u(p)=F_u(p^2)[\not\!p+M_u(p^2)]$
and similarly for the $d$-quark.
The quark-vector-meson couplings are estimated from a standard
quark model analysis of the nucleon-vector-meson coupling.  We find that
$g_{q\rho}\simeq (3m_q/5M_N)g_\rho(1+C_\rho)$ and
$g_{q\omega}\simeq (m_q/M_N)g_\rho(1+C_\omega)\simeq 
(m_q/M_N)g_\rho$.  With quark masses of $\simeq 400$MeV this gives
$g_{q\rho}\simeq 3.9-4.3$ and $g_{q\omega}\simeq 4.8-5.2$ as typical
ranges for these couplings.  Throughout this work we will use
then $g_{q\rho}g_{q\omega}=20.0$.

In the chiral limit, (i.e., with massless quarks and
neglecting electromagnetic and weak effects), the pion decay
constant can be related to the quark propagator through the
Goldberger-Treiman relation (see, e.g., \cite{wkr} and references therein)
to give
\begin{equation}
f_\pi^2=(-12i)\int{d^4\ell\over (2\pi)^4}\; {F(\ell^2)M(\ell^2)\over
\ell^2-M^2(\ell^2)}\; \bigg[ M(\ell^2) - \frac{1}{2}\ell^2\frac{dM}{d\ell^2}
(\ell^2)\bigg]\;.
\label{fpi}
\end{equation}
Factors of three and four follow from colour and spinor traces respectively.
(Note that this differs slightly from the expression in Ref.~\cite{wkr}
which used a dressed axial-vector vertex instead of the appropriate
bare one.  The bare vertex avoids double counting problems).
Similarly, in the chiral limit (see, e.g., \cite{wkr} and references
therein) we obtain an expression for the quark condensate in terms
of the quark propagator
\begin{equation}
\langle\bar qq\rangle\equiv -m_0^3
=-i\int{d^4\ell\over(2\pi)^4}\;{\rm Tr}S(\ell)
=(-12i)\int{d^4\ell\over(2\pi)^4}F(\ell^2)M(\ell^2) \;,
\label{qbarq}
\end{equation}
where the trace is over colour and spinor indices.
We use these two results to constrain the quark propagators 
by fitting them to the experimental values of $f_\pi=93$MeV and
$\langle q\bar q\rangle\equiv m_0^3 =-(225{\rm MeV})^3$,
(i.e, $m_0=225$MeV).

Motivated by the confining quark model from the Dyson-Schwinger
equation study of Ref.~\cite{brw} and the need to avoid the artificial
quark pair-production threshold encountered by GHT in the timelike region
we first consider the following form for the quark propagator [c.f.,
Eq.~(\ref{Sq})]:
\begin{equation}
F(q^2)=f_0 e^{-q^2/\mu^2}\hskip0.5cm{\rm and}\hskip0.5cm M(q^2)=m\;,
\end{equation}
where $f_0$ and $\mu$ are parameters and $m$ is to be chosen as a typical
infrared quark mass.  This propagator is confining and analytic
everywhere in the complex $q^2$ plane (i.e., entire).  It captures the
essential elements of the propagator of Ref.~\cite{brw} and has the
great advantage that with the appropriate choice of gaussian
quark-vector-meson form factors we can use Eq.~(\ref{Pi})
to obtain an analytic expression for the mixing amplitude $\Pi(q^2)$
in both the timelike and spacelike regimes.
In practice it is easier to calculate the expressions for $f_\pi$
and the condensate $m_0$ [i.e., Eqs.~(\ref{fpi},\ref{qbarq})] after first
rotating to Euclidean space.  Evaluating these expressions for
our analytic (confining) propagator gives
\begin{eqnarray}
f_0&=&4\pi^2 m_0^3/3m\mu^4 \nonumber \\
f_\pi^2&=&{3f_0 m^2\mu^2\over 4\pi^2}
\bigg[1\;-\;\bigg(ae^a\int^\infty_a dt\;{e^-t\over t}\bigg)
\bigg]_{a=(m/\mu)^2}\;. 
\label{aparams}
\end{eqnarray}
Hence we can now fix the parameters $f_0$ and $\mu$.
Motivated by the infrared values of the dynamically
generated quark mass in Dyson-Schwinger
equation studies\cite{wkr} we choose here $m=450$MeV for the ``constituent''
quark mass.  Then substituting the expression for
$f_0$ into the expression for $f_\pi^2$
in Eqs.~(\ref{aparams}) lets us solve for $\mu$ to give
$\mu=523.6$MeV and hence we then find
$f_0=4.432\times10^{-6}$MeV$^{-2}$.  Thus all parameters are now
constrained.

Since the $u$-$d$ quark mass difference is commonly estimated
to be $\simeq 4$MeV we use $m_d-m_u=4$MeV throughout this work.
Hence the ``constituent'' masses are $m_u=450$MeV and $m_d=454$MeV
and $f_{0u}=f_0$ and $\mu$ are as determined above.
The simplest and most reasonable assumption is that $\mu$ has the same value
for the $u$ and $d$ quarks, and hence from Eq.~(\ref{aparams})
$f_{0u}/f_{0d}=m_d/m_u$. 
As discussed above in order to facilitate
the integration in Eq.~(\ref{Pi}), we use
$F_{q\rho}(\ell^2)=F_{q\omega}(\ell^2)=\exp(\ell^2/\Lambda^2)$ with
$\Lambda=1$GeV, which is a typically reasonable number for the
form factor cut-off.  We see that these form factors fall off as gaussians
in the spacelike region.
The integral in Eq.~(\ref{Pi})  can be performed easily and one obtains
for $\Pi(q^2)$
\begin{equation}
\Pi(q^2)={3g_{q\rho}g_{q\omega}\over 16\pi^2\lambda^2}
e^{q^2/2\mu^2}\bigg\{\bigg[f_{0u}^2
[-q^2-4m_u^2-(4/\lambda)]
\bigg]
\; - \;\bigg[u\to d\bigg]\bigg\} \;,
\label{pianalytic}
\end{equation}
where $\lambda\equiv 2[(1/\Lambda^2)+(1/\mu^2)]$.
The mixing amplitude has thus been calculated for all timelike and spacelike
$q^2$ for the analytic (confining) case.

We can also numerically evaluate $\Pi(q^2)$ for numerical solutions
of quark Dyson-Schwinger equations.  These solutions are fitted to both
$f_\pi$ and the quark condensate in the usual way\cite{wkr} and in addition
have the correct perturbative asymptotic behaviour in the spacelike
(Euclidean) region.  Unfortunately it is not yet known how to continue
these solutions into the timelike regime in the general case.
We have introduced a 4MeV mass splitting of the infrared mass and have
solved for the $u$ and $d$ propagators in the usual way\cite{wkr}.
The quark-vector-meson couplings ($g_{q\rho}$ and $g_{q\omega}$)
have already been deduced and
for the corresponding form-factors we choose the form used by GHT
for ease of comparison, i.e.,  
$F_{q\rho}(\ell^2) =F_{q\omega}(\ell^2) =1/[1-(\ell^2/\Lambda^2)]$.
We again use $\Lambda=1GeV$.
We can now use Eq.~(\ref{Pi}) to evaluate $\Pi(q^2)$ for a typical
Dyson-Schwinger equation solution in the spacelike region.

Finally, for comparison we repeat the calculation with
the numerical Dyson-Schwinger solution replaced by a free quark
propagator with artificially heavy quarks ($m_u=600$MeV and
$m_d=604$MeV).  This calculation is then identical to that of GHT
except that now there is no troublesome quark threshold between
the spacelike region and the $\omega$-mass point where the
mixing amplitude is actually measured.

In order to understand the importance of the momentum dependence 
of the mixing amplitude on the nucleon-nucleon potential due to vector meson
exchange, we consider the static, central part of this potential. 
Besides a central potential, the exchange of a vector boson gives rise to 
spin-orbit, spin-spin and tensor potentials. Although the spin-dependent 
potentials are the relevant ones for the TRIUMF and IUCF experiments, the 
radial part of these potentials will be directly related to that
of the central one. The coordinate space $NN$ potential due to
$\rho\omega$-mixing, $V(r)$, is the Fourier transform of $V(q)$ in
Eq.~(\ref{Vq}), where $q^0=0$ and where we here define $q\equiv|\vec q\,|$
\begin{eqnarray}
V(r) &=& \frac{1}{2\pi^2r} \int^{\infty}_{0} dq\; q\; \sin(qr)V(q)
\nonumber \\
&=& \frac{1}{2\pi^2r} \int^{\infty}_{0} dq\; q\; \sin(qr)
\frac{-g_\rho g_\omega F_{N\rho}(-q^2) F_{N\omega}(-q^2)
\Pi (- q^{2})}{(q^{2} + m^{2}_{\rho}) (q^{2} +
m^{2}_{\omega})}\;.
\label{Vr}
\end{eqnarray}


In Fig.~1 we show the momentum-dependence of the $\rho^0$-$\omega$
mixing amplitude for the three quark calculations described here,
i.e., the confining (analytic), Dyson-Schwinger equation (dse), and
heavy-free (free) cases.  Also shown for comparison are recent hadronic
calculations\cite{pw}, where an essentially parameter free calculation
was made using an $N\bar N$ loop and the $n$-$p$ mass difference. The two
hadron-model curves differ in that one (hadron\_ff) has the usual Bonn
form factor applied at the vertices of the nucleon loop in the spaceklike 
region as well as at the meson-nucleon vertices.
We see that all calculations appear capable of fitting the data with
typical parameters and that the momentum dependence is remarkably
similar in each case.  We see strong momentum-dependence of the mixing
amplitude and a change in sign in the vicinity of $q^2=0$.  This is
also what was seen in the recent QCD sum rules study\cite{hhmk},
although there $\Pi(q^2)$ was seen to change sign in the timelike region
at $q^2\simeq0.25-0.40$GeV$^2$ rather than the spacelike region. 

In Fig.~2 we show the momentum-space potential $V(q)$ given in
Eq.~(\ref{Vq}) for the various models.  Also
for comparison we show the result
obtained when the usual assumption of momentum-independence of the mixing
amplitude is made.  In Fig.~3a) we show the corresponding coordinate-space
potentials $V(r)$ given by Eq.~(\ref{Vr}) for each of these cases.  Fig.~3b)
has an enhanced vertical scale in order to show the location of
nodes in the potential $V(r)$.  We see that all calculations predict
strong suppression of the $\rho^0$-$\omega$ mixing
potential compared with that resulting from the usual momentum-independent
assumption for the mixing amplitude.   
We see that the Dyson-Schwinger propagator and the hadronic
models predict the opposite sign at small $r$ and have nodes
between 0.5 and 0.8fm, whereas the heavy free quark
and confining (analytic) quark cases have the same sign as the usual assumed
potential, but are very strongly suppressed.
It is interesting to note that the heavy free quark case presented here
is identical to the case studied by GHT with the exception of having
heavier (i.e., 600MeV) quarks. We see that this increase in mass
has removed the node in $V(r)$ for the free propagator case.
Given the variety of approaches to the calculations,
the results are remarkably similar.

The possible implications for our understanding of CSB are far-reaching.
For example, in the case of the $\rho$-$\omega$ mixing
contribution to the class IV CSB in $n$-$p$ elastic
scattering\cite{triumf}-\cite{bw} there is a 
competition between large-$r$
fall-off in the potential and short-distance suppression of the
distorted $N$-$N$ wavefunction.  The result is that the conventional
contribution peaks around 0.9fm.  It is interesting that this is the region
where nodes occur.  All models predict at least strong suppression in this
region.
It is clear that the previously assumed theoretical understanding of
charge-symmetry breaking phenomena is now questionable.
The models and treatments to date appear to be in overall agreement.
The crucial question is whether these conclusions will survive future,
more elaborate examinations.

\begin{center}
  Acknowledgements
\end{center}

We are thankful to K. Yazaki for for a helpful discussion. 
This work was supported by the Australian Research Council (AWT and AGW)
and CNPq (Brasil) (GK).
AGW and GK were also partially supported by the U.S. Department of Energy 
through Contract No. DE-FG05-86ER40273 and by the Florida State University
Supercomputer Computations Research Institute which is partially funded by
the Department of Energy through Contract No. DE-FC05-85ER250000.

\newpage

\noindent
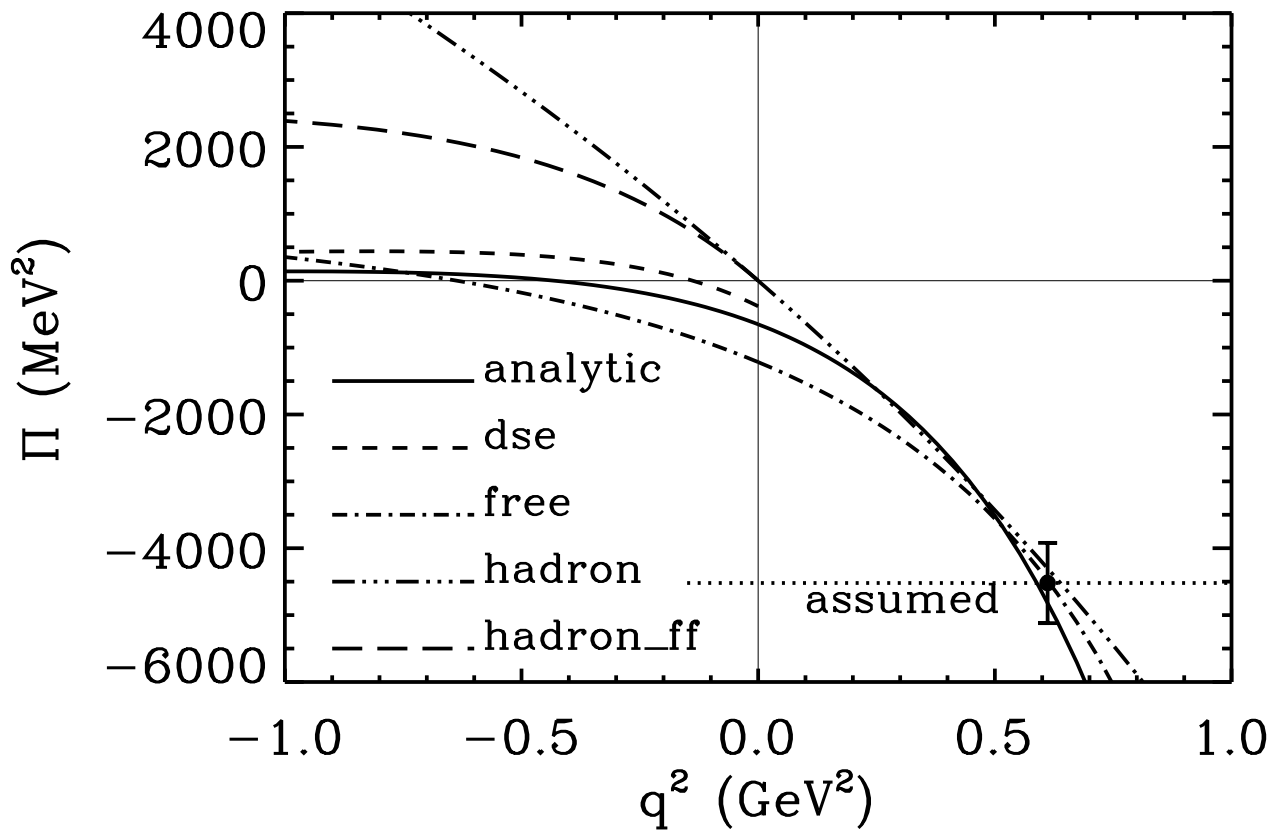
{\bf Fig. 1} The $\rho$-$\omega$ mixing amplitudes calculated
in a variety of models, (see text).  Also shown for comparison is
the experimentally extracted amplitude at the $\omega$-mass-shell 
point and the usual assumed $q^2$-independent behaviour. 
\vskip12pt
\noindent
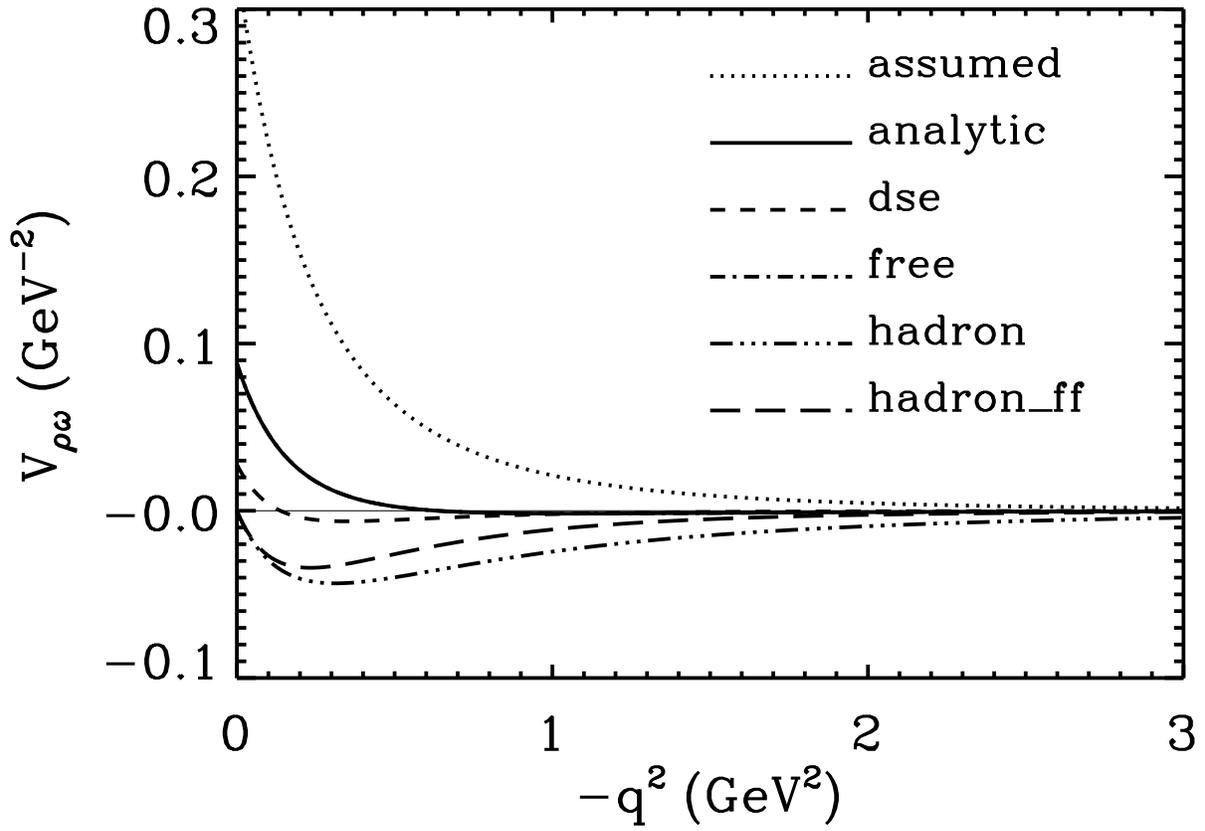
{\bf Fig. 2}  The $\rho$-$\omega$
momentum space potentials, $V(q)$, for the
mixing amplitudes shown in Fig.~1.  $V(q)$ is defined in the text.
\vskip12pt
\noindent
{\bf Fig. 3}  The $\rho$-$\omega$ coordinate space
potentials, $V(r)$, for the
mixing amplitudes shown in Fig.~1.  $V(r)$ is defined in the text.
We show the full range of the potentials in a), while in b) an enhanced
vertical scale is used to show in detail how the potential changes sign
in some cases.

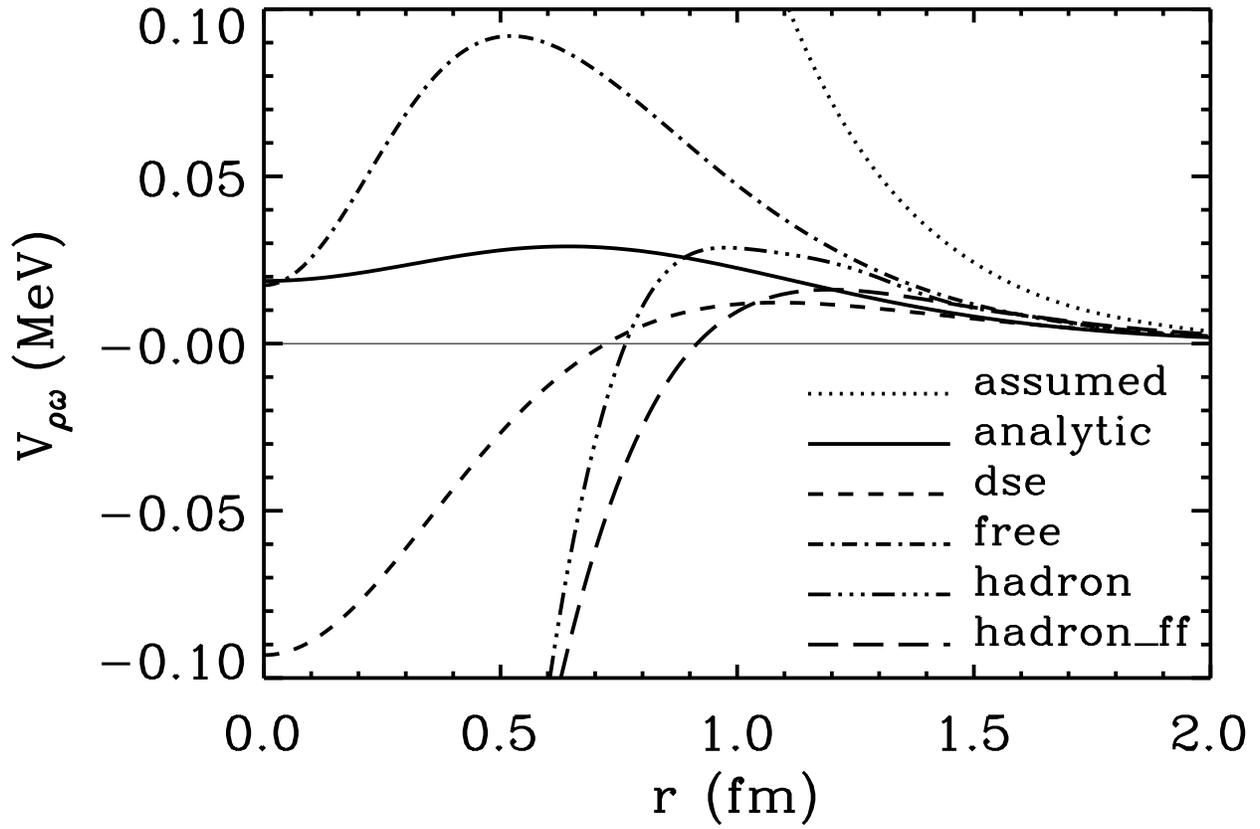
\end{document}